# A Simple and Reliable Formula for Assessment of Maximum Volumetric Productivities in Photobioreactors

*Jean-François CORNET* and Claude-Gilles DUSSAP*

Clermont Université - Laboratoire de Génie Chimique et Biochimique, Bât. Polytech.

24, avenue des Landais – BP 206

63174 AUBIERE Cedex – France

Phone : (33) *4-73-40-50-56

Fax : (33) *4-73-40-78-29

J-Francois.CORNET@univ-bpclermont.fr

**ABSTRACT**

This paper establishes and discusses the consistency and the range of applicability of a simple, but general and predictive analytical formula, enabling to easily assess the maximum volumetric biomass growth rates (the productivities) in several kinds of photobioreactors with more or less 15% of deviation. Experimental validations are performed on photobioreactors of very different conceptions and designs, cultivating the cyanobacterium *Arthrospira platensis*, on a wide range of volumes and hemispherical incident light fluxes. The practical usefulness of the proposed formula is demonstrated by the fact that it appears completely independent of the characteristics of the material phase (as the type of reactor, the kind of mixing, the biomass concentration…), according to the first principle of thermodynamics and to the Gauss-Ostrogradsky theorem. Its ability to give the maximum (only) kinetic performance of photobioreactors cultivating many different photoautotrophic strains (cyanobacteria, green algae, eukaryotic microalgae) is theoretically discussed but experimental results are reported to a future work of the authors or to any other contribution arising from the scientific community working in the field of photobioreactor engineering and potentially interested by this approach.

# 1- Introduction and Objectives

Modeling, design and scale-up of photobioreactors (PBR) is a research field of fast increasing importance because the consciousness that photosynthesis will play a crucial role in the future, both in the mitigation of the $CO_2$ emissions and in the sustainable and renewable synthesis of chemicals (biofuels, raw molecules for chemistry or high valuable products), is progressively imposed to the population all around the world (*1-3*). In this prospect, the cultivation of photosynthetic microorganisms in PBR enabling higher productivities, a rigorous control of minerals and an optimized management of water fluxes is the only way to avoid an unacceptable competition with plant crop cultivation for food.

Because of the specific requirement of light (sunlight or artificial light) as a main energy source for the PBR functioning, a great variety of concepts exists in the literature (*4*, *5*), even if, from a chemical engineering point of view, they can just be subdivided in completely stirred tank reactors (CSTR) and plug flow tubular reactors (PFTR). The former are generally used for applications with artificial light and present a strong advantage to operate with a spatial homogeneous biomass concentration, enabling a fine control of the field of radiation inside the reactor. At the opposite, PFTR are often preferred for sunlight applications as they seem more adapted to large reactors designed for outdoor cultivation, except if the recent concept of internally lighting PBR is envisaged (*6-8*).

Whatever the variety of existing concepts and their intrinsic complexity, engineering purposes require to establish simple but predictive formula for the assessment of kinetic and energetic performances of PBR, mainly useful for design and scale-up operations, or for comparison of different systems between them, on a Cartesian basis, at the opposite of a discourse on the famous effect of the "biological diversity". Moreover, the idea that more complicated is the theoretical demonstration of a result, less is its practical usefulness, or at the opposite, that only simple demonstrations and formula are of any practical usefulness for potential applications, is known as the Hartmanis' principle and has been partly demonstrated recently using the Kolmogorov complexity theory (*9*). The aim of the present paper is to demonstrate that, even if a complete PBR understanding and modeling remains rigorously a very complex problem with numerous intricate levels of description (see hereafter), it is however possible, according to the Hartmanis' principle, to easily establish a simple but general and reliable formula for the assessment of the maximum volumetric biomass productivities. The fundamental interest of such a predictive analytical formula, if it can be established, is that it gives always definitive and general responses and solutions whereas complex and full numerical simulations only depict in more details a particular situation. As stated by the Hartmanis' principle, the authors hope that the simple and easy-to-use relation proposed in this paper will be of great practical interest to clarify and drive the future developments in the field of PBR engineering at two levels. First, the rigorous demonstration itself gives enlightenments on the key variables enabling to properly formulate the coupling between light transfer and kinetic rates in PBR, clarifying by the way many important points which often lead to misinterpretations in the specialized literature. Second, the final concise formula establishes the respective importance of each main variable in the assessment of kinetic performances of PBR with associated lumped parameters, which is of considerable interest in the perspective of designing new efficient PBR for future applications.

Clearly, it is not envisaged in this paper to take under consideration all the factors influencing the biomass growth rate in a PBR (*5*). Stating that the main limiting engineering factor is the light distribution and availability in the reactor (*10-16*), only the effect of the radiant light transfer on modeling and its consequences on the mean biomass volumetric productivities observed in PBR will be investigated in the following. All the other important parameters (culture medium, temperature, pH, dissolved gases…) will be assumed to be controlled in optimal conditions along this work. The useful and reliable formula will be first established in the framework of photoautotrophic CSTR photobioreactors, but a possible extension of its domain of validity in calculating maximum kinetic performances for PFTR photobioreactors will be then discussed. In all cases, a special attention will be paid on the energetic consistency of the obtained theoretical result, as a proof of the global coherence of the proposed approach. The experimental validations of the final formula will be envisaged on eight artificially lightened PBR of very different conceptions and designs with volumes varying on three order of magnitude and hemispherical incident photon flux densities (PFD) varying on two orders of magnitude. At this stage however, only one model of photoautotrophic microorganism, the cyanobacterium *Arthrospira platensis*, will be used to prove the validity of the theoretical approach developed in the paper.

## 2- Materials and Methods

For all the experiments, the microorganism used was *Arthrospira* (*Spirulina*) *platensis* PCC 8005 (Institut Pasteur, Paris, France). It was grown axenically in Zarrouk medium modified by Cornet (*17*) or by Cogne (*18*) for trace elements, depending of the experiment (but always using nitrate as nitrogen source). The pH (between 8 and 10) and the temperature (35-36°C) were controlled at their optimal values for growth. Results for eight different kinds of artificially lightened PBR (CSTR) were reported in this study with working volumes varying between 0.1 and 77 L and incident hemispherical photon fluxes ranging between 30 and 1600 $\mu mol.m^{-2}.s^{-1}$ (PAR, using multi-points measurements with a LI-COR cosine quantum sensor LI-190SA and confirmed by actinometry in case of PBR marked with a star in the following). The main parameters influencing the biomass productivities (geometry, characteristics of the lighting system) and the culture conditions (mixing, pH and temperature control) are depicted hereafter for each kind of PBR.

- PBR 1: rectangular PBR of 4 L working volume illuminated by one side with 5 fluorescent tubes (Philips, white industry, 20W in the spectral range [380-750 nm]) and agitated by bubbling air-$CO_2$ (2%) at a flow rate of 0.02 vvm (enabling to control the pH) plus magnetic stirring. Temperature was controlled at the level of the room (culture chamber). The path length of the reactor was 8 cm giving a specific illuminated area $a_{light}$ of 12.5 $m^{-1}$ without dark fraction ($f_d = 0$).

- PBR 2: cylindrical stirred tank reactor (Applikon) of 5 L working volume illuminated by one side with two halogen projectors (Philips, 500 W in the spectral range [350-1100 nm]) and agitated by two rushton turbines. Temperature was controlled with an internal exchanger fed with an external thermal bath and pH was regulated using 1 mol/L $H_2SO_4$. The diameter of the reactor was 16 cm giving a specific illuminated area $a_{light}$ of 12.5 $m^{-1}$ without dark fraction ($f_d = 0$). Special attention must be paid in this case that the mean incident hemispherical PFD appearing in Table 2 is given in front of the reactor and must be cosine averaged on half of the cylinder (divided by 2) in order to use in the final eq. (22).

- PBR 3*: cylindrical stirred tank reactor (Applikon) of 5 L working volume radially illuminated by 55 halogen lamps (Claude BAB 38°, 20 W, 12 V in the spectral range [350-1100 nm]) and agitated by two rushton turbines. Temperature was controlled with an internal exchanger fed with an external thermal bath and pH was regulated using 1 mol/L $H_2SO_4$. The diameter of the reactor was 16 cm giving a specific illuminated area $a_{light}$ of 25 $m^{-1}$ without dark fraction ($f_d = 0$).

- PBR 4: cylindrical air-lift reactor (Bioengineering) of 7 L working volume radially illuminated by 48 halogen lamps (Sylvania BAB 38°, 20W, 12 V in the spectral range [350-1100 nm]) and agitated by bubbling air-$CO_2$ at a flow rate of 1 vvm. Temperature was controlled with an internal exchanger fed with an external thermal bath and pH was regulated by adjusting automatically the $CO_2$ mole fraction in the input gas. The diameter of the reactor was 10 cm giving a specific illuminated area $a_{light}$ of 40 $m^{-1}$ with a dark volume fraction $f_d = 0.48$.

- PBR 5*: Oblate cylindrical membrane reactor of 0.106 L working volume lightened by the height with one halogen lamp (Claude BAB 38°, 20 W, 12 V in the spectral range [350-1100 nm]) and magnetically stirred. Temperature was controlled with an exchanger fed with an external thermal bath and pH was regulated by $CO_2$ pulses in the gas phase when necessary. The path length of the reactor was 2.3 cm giving a specific illuminated area $a_{light}$ of 43.5 $m^{-1}$ without dark fraction ($f_d = 0$).

- PBR 6: double cylindrical (riser and down comer) air-lift reactor (Bioengineering) of 77 L working volume radially illuminated by 350 halogen lamps (Sylvania BAB 38°, 20W, 12 V in the spectral range [350-1100 nm]) and agitated by bubbling air-$CO_2$ at a flow rate of 1 vvm. Temperature was controlled with an internal exchanger fed with an external thermal bath and pH was regulated by adjusting automatically the $CO_2$ mole fraction in the input gas. The diameter of each cylinder was 15 cm giving a specific illuminated area $a_{light}$ of 26.7 $m^{-1}$ with a dark volume fraction $f_d = 0.33$.

- PBR 7: cylindrical reactor with a centrifugation field generated by 4 paddles (300 rpm) and radially illuminated by 44 fluorescent white tubes (Philips, 18 W, in the spectral range [380-750 nm]). The total volume of the reactor was 13.2 L (total diameter of 22 cm) but working volume was only 6 L (biofilm thickness of 3 cm at the wall). Temperature was regulated using fans and convective air in order to remove the heat of the lamps and pH was controlled with $H_2SO_4$, 1 mol/L. The corresponding specific illuminated area $a_{light}$ was 40 $m^{-1}$ without dark volume fraction $f_d = 0$.

- PBR 8*: rectangular PBR of 0.5 L working volume illuminated by one side with a panel of 180 red LED (Agilent, 38mW - 16000 mcd – 8°, in the spectral range [635-655 nm]) and agitated by bubbling air-$CO_2$ at a flow rate of 0.02 vvm (with a regulated $CO_2$ mole fraction enabling to control the pH) plus magnetic stirring. Temperature was controlled using a thermostatic bath and external heat exchanger. The path length of the reactor was 4 cm giving a specific illuminated area $a_{light}$ of 25 $m^{-1}$ without dark fraction ($f_d = 0$).

The dry biomass concentration (DM) was systematically determined by weighting filtered samples (Millipore, 0.45 µm) preliminary dried at 110°C during 24 hours. The dry mass productivities were determined according to the material balance for biomass (eq. 24), i.e. measuring the output flow rate and (light-limited) biomass concentration over at least six stable residence times for continuous cultures, or determining the slope of the (light-limited) biomass concentration time course for batch cultures.

# 3- Theoretical Considerations

## *3.1- The Complexity of a Predictive Knowledge Model for PBR*

As already explained by the authors (*19*), a general thorough understanding and knowledge modeling of PBR is a very difficult task requiring to manage a seven dimensional Euclidean space in which the relaxation times of the involved kinetic phenomena overlap twelve order of magnitude. This relies first on a fine description of the spectral radiant light transfer inside the reactor for any geometry and boundaries conditions, because in the general case, formulating the local kinetic coupling requires to accurately determine the local radiant power density, considering the PBR as a scattering and participative medium. This work is only rigorously feasible using the spectral (Boltzmann-type) quasi steady-state radiative transfer equation (RTE) inside the PBR, leading to solve (*13*, *20*):

$$\nabla \cdot \left(\mathbf{\Omega} I_\lambda\right) = -\left(Ea_\lambda + Es_\lambda\right) C_X I_\lambda + \frac{Es_\lambda C_X}{4\pi} \iint_{4\pi} I_\lambda \, p_\lambda\left(\mathbf{\Omega}, \mathbf{\Omega}'\right) d\Omega' \quad (1)$$

The spectral absorption $Ea_\lambda$ and scattering $Es_\lambda$ coefficients with the phase function for scattering $p_\lambda\left(\mathbf{\Omega}, \mathbf{\Omega}'\right)$ (i.e. the so-called radiative properties for the medium) appearing in eq. (1) can be obtained using sophisticated experimental methods (*21*), but recently, the authors have proposed an entirely predictive methodology, using the generalized Lorenz-Mie theory and requiring only to know some basic information as a model of shape, a size distribution and the pigment content for the microorganism (*7*, *22*, *23*). Finally, eq. (1) must be solved for a given geometry of PBR and boundary illuminating conditions, using one-dimensional methods in simple cases like modulated two-flux and discrete ordinate methods (*7*), or more sophisticated Monte Carlo (*13*, *14*) and finite element methods (*20*) in the general case. The spectral field of irradiance $G_\lambda$ is then known everywhere inside the reactor as a preliminary stage in formulating the kinetic coupling, by averaging over all the wavelength in the PAR both the field of irradiance $G = \int_\lambda \iint_{4\pi} I_\lambda \, d\lambda \, d\Omega$ and the field of radiant light power density $\mathcal{A} = \int_\lambda \iint_{4\pi} Ea_\lambda \, C_X \, I_\lambda \, d\lambda \, d\Omega$.

The former quantity enables to define the working illuminated fraction $\gamma = V_\ell / V$ inside the reactor (from the localization of the compensation point for photosynthesis in relation to the irradiance $G_C$), corresponding to a volume in which photosynthetic growth occurs, and then to write the mean volumetric growth rate in the completely stirred PBR as a result of two different averaged metabolic zones (*11*, *12*, *24*):

$$< r_X >= \gamma \frac{1}{V_\ell} \iiint_{V_\ell} r_{X,\ell} \, dV + (1-\gamma) \frac{1}{V_d} \iiint_{V_d} r_{X,d} \, dV \qquad (2)$$

For cyanobacteria, and specially for *A. platensis*, having common electron carriers chains for photosynthesis and respiration, it is well established (*7*) that a first analysis of the relaxation time necessary to switch the metabolism from photosynthesis to respiration leads to consider in any case that $r_{X,d} = 0$ in the dark volume $V_d$, even in partially lightened PBR with low values of the $\gamma$ fraction. As a consequence, the maximum and constant volumetric growth rate for cyanobacteria is reached as soon as the $\gamma$ fraction is lower or equaled to 1, corresponding to the well-known physical limitation by light transfer in PBR (all the incident photons are absorbed and used in the vessel). Evidently, as it will be discussed later in the paper, this assumption is not true for eukaryotic microorganisms as microalgae for which photosynthesis and respiration operate separately in chloroplasts and mitochondria and nor for photoheterotrophic bacteria (*19*).

The latter quantity $\mathcal{A}$, enables to formulate the local kinetic coupling in the illuminated volume, leading to the final knowledge expression for $r_{X,\ell}$. It is well-known indeed that this coupling rigorously relies on the local radiant light power density $\mathcal{A}$ (*13*), defining properly the energetic yield $\rho$ and the mass quantum yield $\bar{\phi}$ of conversion, as already proposed by the authors (*19*, *25*):

$$r_{X,\ell} = \rho \bar{\phi}\, \mathcal{A} = \rho \bar{\phi}\, Ea\, C_X\, G \qquad (3)$$

As recently explained (*7*, *23*), this partition is a consequence of first principle analyses of the excitation energy transfer in the photosynthetic antenna toward the reaction center and then of the so-called Z-scheme for photosynthesis leading to the synthesis of ATP and reducing power $NADPH_2$. The energetic yield $\rho$ represents the dissipative part of the photonic energy absorbed by the antenna. This local value decreases with the local irradiance from a convenient relationship obtained on the basis of an approximation of a theoretical quantum physics study of the excitation transfer mechanisms in antenna (*27*):

$$\rho \cong \rho_M \frac{1}{1 + \dfrac{G}{K}} = \rho_M \frac{K}{K + G} \qquad (4)$$

in which $\rho_M$ is the maximum value of the yield, obtained when the system operates in the thermodynamic optimal conditions, at the irradiance of compensation $G_C$ (the irradiance $G$ becomes negligible in regard to the half-saturation constant $K$). It must be pointed out at this stage that, even if eq. (4) remains a representation model, it gives some theoretical background to the well-known but sometimes discussed hyperbolic behavior of the photosynthesis in regard to the local irradiance.

At the opposite, the mass quantum yield $\bar{\phi}$ is a mean value in time (as indicated by the bar), at a time scale corresponding to the pseudo-steady state hypothesis for the products of the Z-scheme for classical photosynthesis like ATP and $NADPH_2$ (one to some minutes). It corresponds to the conservative part of the photonic energy absorbed in the reactor (the part of the photonic processes leading to a charge separation at the reaction center), and has been demonstrated to be consequently a mean value in space (*7*) for the general case of well-mixed photobioreactors. It can then be easily deduced from the true mole quantum yield $\overline{\phi}'$ by just using the mean C-molar mass of the produced biomass in the reactor $M_X$:

$$\overline{\phi} = M_X \overline{\phi}' \quad (5)$$

As discussed hereafter, this true mole quantum yield $\overline{\phi}'$ may be obtained by a stoichiometric analysis of the Z-scheme, taking into account the cyclic photophosphorylation from the well-known P/2e$^-$ ratio (*12*, *23*), completing then the proposed knowledge model for the kinetic coupling, using finally from eqs. (3-5) the local equation:

$$r_{X,\ell} = \rho_M \, K \, \overline{\phi} \, Ea \, C_X \, \frac{G}{K+G} \quad (6)$$

## *3.2- Reification for the Kinetic Parameters Calculation*

Clearly, the proposed approach for the local kinetic coupling is completely predictive, only if all the parameters and variables involved in eq. (6) can be determined *a priori* by a theoretical analysis. As already explained above with the help of the relevant literature, it is the case for all the data involved in the physical phenomenon of radiant light transfer in scattering and participating dense media, as the spectral or mean field of irradiance *G*, and the mean absorption coefficient for the biomass *Ea*.

Among the three other kinetic parameters requiring a reification $\rho_M$, $\overline{\phi}$ and *K*, the maximum energetic yield $\rho_M$ will be first discussed in the following. It corresponds at the maximum thermodynamic efficiency, or the minimum losses, for the photonic conversion and excitation transfers in the photosynthetic units or photosynthetic domains (*26*) towards the reaction centers, leading finally to a charge separation process. Even if the maximum efficiency for radiant energy conversion processes has been strongly debated in the past and clarified by Bejan (*27*, *28*), we can retain here the most simple approach proposed by Jeter and used for photosynthesis from a long time (*29*). These authors just consider an extension of the Carnot formula for the maximum radiation conversion (*27*):

$$\rho_M = 1 - \frac{T}{T_R} \quad (7)$$

in which the temperature for the radiation is given by the ideal black body formula of Planck, considering for the spectral intensity $I_{C,\bar{\lambda}}$ in thermodynamic optimal conditions, the value at the compensation point for the mean considered wavelength $\bar{\lambda}$ in the reactor:

$$T_R = \frac{\mathrm{h}\, n_m\, \mathrm{c}}{\mathrm{k}\,\bar{\lambda} \ln\left[1 + \dfrac{2\,\mathrm{h}\, n_m^2 \mathrm{c}^2}{I_{C,\bar{\lambda}}\,\bar{\lambda}^5}\right]} \quad (8)$$

It is obvious that $I_{C,\bar{\lambda}}$ depends on the value of the irradiance of compensation $G_C$, on the angular nature of the radiation field for the radiant intensity calculation, and on the spectral nature of the field for the calculation of $\bar{\lambda}$. Nevertheless, evaluating $\rho_M$ from eqs. (7-8) in a wide range of previous conditions, both for prokaryotic and eukaryotic microorganisms having different optimal temperatures of functioning, leads generally to numerical values ranging between 0.76 and 0.82 (in the PAR). This enables to retain in first approximation the constant value $\rho_M = 0.8$ with less than 10% deviation for any photosynthetic microorganism considered in this study. It should be mentioned here that this value is in very good agreement with the theoretical ones determined by a thorough analysis of the excitations transport in antenna on a quantum physics basis (*26*), demonstrating that, around the compensation point, photosynthesis operates close to the optimum thermodynamic conditions.

The true mole quantum yield $\bar{\phi}'$, enabling to easily derive the mass quantum yield $\bar{\phi}$ (eq. 5) can be also determined from a stoichiometric analysis of the Z-scheme for photosynthesis (*23*). This requires establishing a structured stoichiometry for a given photoautotrophic microorganism (*30*) as it has been done for *A. platensis* for reasonably low hemispherical incident photon fluxes (*12*):

$$\begin{aligned} & CO_2 + 1.463\,H_2O + 0.173\,HNO_3 + 0.006\,H_2SO_4 + 3.544\,ATP + 2.779\,(NADPH,H^+) \\ & \xrightarrow{J_X} CH_{1.575}O_{0.459}N_{0.173}S_{0.006}P_{0.006} + 3.538\,P_i + 3.544\,ADP + 2.779\,NADP^+ \end{aligned} \quad (9)$$

Eq. (9) is associated to the corresponding Z-scheme stoichiometric equation:

$$\begin{aligned} 2.779\,NADP^+ + 2.779\,H_2O &\xrightarrow{J_{COF}} 2.779\,(NADPH,H^+) + 1.389\,O_2 \\ 3.544\,(ADP + P_i) &\xrightarrow{J_{ATP}} 3.544\,ATP + 3.544\,H_2O \end{aligned} \quad (10)$$

eliminating by summation the cofactors and the ATP syntheses and demonstrating that, if the N-source for photosynthesis was nitrate, the resulting photosynthetic quotient $Q_P$ ($Q_P = J_{O_2} / J_{CO_2}$) is roughly 1.4. Obviously, eq. (10) provides the mean value of the P/2e$^-$ ratio for the considered conditions:

$$P/2e^- = \frac{J_{ATP}}{J_{COF}} = 1.275 \quad (11)$$

Finally, the calculation of the true mole quantum yield for the Z-scheme, corresponding to the conservative part of the photonic energy conversion, and only P/2e$^-$ dependent is straightforward (from eqs. 9-11):

$$\bar{\phi}' = \frac{1}{2\upsilon_{NADPH,H^+-X}(1 + P/2e^-)} \cong 7.8 \times 10^{-8}\ \mathrm{mol_X.\mu mol_{h\upsilon}^{-1}} \quad (12)$$

Of course, this value seems linked to the macromolecular composition of *A. platensis* via eq. (9) and restricted to the conditions of low incident PFD (eqs. 10-11), because it has been demonstrated that the P/2e$^-$ ratio varied significantly with the hemispherical incident photon flux (*7*, *12*). Nevertheless, from a detailed theoretical analysis of the Z-scheme functioning, using the linear thermodynamics of irreversible processes (requiring lengthy and tedious developments remaining out of the scope of this paper), the authors proved (*7*) that the numerical value (eq. 12) appeared quasi-constant in all the domain of incoming PFD onto the PBR, because the increase of the P/2e$^-$ ratio was counterbalanced by a decrease in the stoichiometric coefficient $\upsilon_{NADPH,H^+-X}$ related to the global formula of the produced biomass quality. Additionally, the theoretical approach used being general (at the thylacoïd level) and completely independent of any given microorganism, the numerical value determined by eq. (12) can thus be considered as a quasi universal result (within more or less 10% of deviation) for photosynthesis as long as nitrate ions are used for the N-source and if all the physiological conditions are optimal for growth (pH, temperature). Conversely, the conversion from mole to mass quantum yield $\bar{\phi}$ from eq. (5) can be questionable if the quality of the biomass varies with lighting conditions of the PBR and with the considered microorganism. If it is rigorously true that the macromolecular composition is different for each microorganism and culture conditions, except for severe substrate limitations which are not considered in this paper, it has been well-documented again (*30*) that the resulting C-molar mass $M_X$ for the biomass rarely exceed 10% of deviation. Taking for photosynthetic microorganisms the mean value $M_X = 0.024\,\mathrm{kg.mol^{-1}}$ leads definitively to the general value:

$$\bar{\phi} \cong 1.85 \times 10^{-9}\ \mathrm{kg_X.\mu mol^{-1}_{h\upsilon}} \quad (\pm 15\%) \qquad (13)$$

As briefly sketched above, in the optimal conditions for growth (only light-limited cultures, no mineral limitations) this quantum yield is mainly affected by the corresponding N-source. For example, the previous global analysis could be perform again considering ammonia ($NH_4^+$) as unique N-source for growth (an interesting case for many photosynthetic microorganisms). In this case, the photosynthetic quotient of the stoichiometric equations (9-10) becomes roughly $Q_P$ = 1.1, that is $\upsilon_{NADPH,H^+-X} \cong 2.2$, and using eq. (12) with the same C-molar mass $M_X$ leads to the quantum yields values:

$$\begin{aligned} \bar{\phi}' &\cong 1.0 \times 10^{-7}\ \mathrm{mol_X.\mu mol^{-1}_{h\upsilon}} \\ \bar{\phi} &\cong 2.4 \times 10^{-9}\ \mathrm{kg_X.\mu mol^{-1}_{h\upsilon}} \end{aligned} \qquad (14)$$

This interesting result demonstrates that the efficiency of photosynthesis is 25% higher if ammonia is used instead of nitrate as nitrogen source, which was a well-known phenomenon on aerobic microorganisms as previously calculated by Roels (*30*). Finally, it is clear again that these proposed optimal values (eqs. 12-14) of quantum yields do not take into account specific conditions of some extremophilic microorganisms cultivated in (for example) highly saline media or at the opposite very low $CO_2$ partial pressures and needing to dissipate important amount of ATP (leading to considerably decrease the previous quantum yields) to maintain a given level of intracellular concentrations or osmotic pressure. This is in any case a limitation of this paper, the aim of which being to calculate the maximum productivities of photobioreactors only.

The last kinetic parameter, the half-saturation constant *K* must be discussed now. Because eq. (4) is only an approximated relation, there is today no mean to have a predictive method giving a theoretical value to *K* as a function of the molecular architecture of the photosynthetic antenna. *K* appears then as the only parameter of the model to be identified and this is indeed a specific value for a given microorganism. Its calculation may be easily done however using for example independent measurements of $O_2$ evolution as a function of the irradiance (if this latter can be rigorously quantified). Once a time, if conducting such experiments on different kinds of photosynthetic microorganisms (prokaryotic or eukaryotic), leads to the determination of surprisingly close values within a range of +/-30%. Nevertheless, such variations indicate that an experimental determination for each considered microorganism can significantly increase the accuracy of the proposed approach. The corresponding value determined for *A. platensis* used in this study has been already widely reported (*12*, *23*) to be:

$$K = 90\ \mu\text{mol}_{h\upsilon}.\text{m}^{-2}.\text{s}^{-1} \quad (15)$$

## *3.3- Problem Reduction in a Particularly Simple Case*

In order to establish and discuss the physical meaning of the final formula, a very simple case (but often encountered in the field of artificially lighting PBR) is first examined. We consider a completely stirred and sufficiently thin rectangular photobioreactor artificially illuminated by one side with a quasi-homogeneous and collimated hemispherical incident photon flux density $q_\cap$. Such geometry enables to make the assumption of one-dimensional light transfer attenuation along a given z-axis, and then to use the simplified form of the RTE in the calculation of the field of irradiance (*21*, *22*), with mean spectrally averaged data on the PAR (absorption and scattering coefficients, phase function, irradiances and incident hemispherical PFD). The possible deviations linked to this assumption have been already discussed (*22*) and do not exceed the degree of approximation envisaged in the following. Unfortunately, the corresponding analytical solution already proposed by the authors (*22*, *23*) is still much more complicated to have an analytical treatment of the kinetics coupling using the integral eq. (2). A new physically consistent approximation of the one-dimensional irradiance profile, deduced from the rigorous two-flux approach and using only one attenuation term is needed. It has been previously established (*11*) that, for a collimated incident hemispherical PFD, the best energetically-consistent approximation was in the form:

$$\frac{G_z}{q_\cap} \cong \exp\left[-\frac{(1+\alpha)}{2\alpha} Ea\, C_X\ z\right] \quad (16)$$

In this equation, $\alpha = \sqrt{Ea/(Ea + 2bEs)}$ is the linear scattering modulus (*22*) defining the so-called backscattered fraction $b = \frac{1}{2}\int_{\pi/2}^{\pi} p\ (\theta, \theta')\sin\theta\ d\theta$ and enabling to correct in a very first approximation the Bouguer's law from scattering. Clearly, this equation is too simple to accurately describe the radiant light profile inside the culture medium and then to give actual local volumetric growth rates in the general case; nevertheless, as it will be demonstrated in the following, it just contains the physical information needed for the objectives of the paper (not more and not less).

Assuming then a coupling formulation for cyanobacteria (a new restriction) leading to take (as previously discussed) $r_{X,d} = 0$ in eq. (2) and with the help of eq. (6) for $r_{X,\ell}$, it becomes easy to perform the resulting spatial integral (eq. 2) over the illuminated volume $V_\ell$ of the CSTR to obtain (assuming that the irradiance of compensation $G_C$ was always sufficiently low to verify $1+G_C/K \cong 1$):

$$< r_X >_{\max} = \rho_M \bar{\phi}\, K \frac{2\alpha}{1+\alpha} \frac{1}{L} \ln\left[1+\frac{q_\cap}{K}\right] \quad (17)$$

This surprisingly simple formula gives a very good approximation of the maximum volumetric growth rate $< r_X >_{\max}$ for any cyanobacteria cultivated in one-dimensional rectangular PBR illuminated by one side. As explained above in this case, it appears completely independent of the illuminated fraction $\gamma$, because the maximum kinetic performances for cyanobacteria (for which respiration is inhibited by light) are reached as soon as the physical limitation by light is established ($\gamma$ is just equaled to one), leading to constant volumetric growth rates for any lower values of $\gamma \le 1$, and then to the observed linear time courses for biomass concentration in batch experiments. To understand the physical meaning of this formula, it must be first noticed that the term $1/L$ corresponds, for the considered case of rectangular reactor illuminated by one side, to the specific illuminated area $a_{light}$ defined from the total working volume of the reactor $V$ by:

$$a_{light} = \frac{S_{light}}{V} \quad (18)$$

Thus, eq. (17) demonstrates that the maximum volumetric biomass growth rate which can be reached in the PBR is mainly dependent on both this specific illuminated area $a_{light}$ and the hemispherical incident PFD $q_\cap$, once the kinetic parameters for the coupling are fixed ($\rho_M, \bar{\phi}, K$). This is the consequence of the physical light transfer limitation, which is (by analogy with the well-known physical limitation by the dissolved oxygen transfer rate) characterized mainly by a geometrical factor ($a_{light}$) and a "force" factor ($q_\cap$). At the opposite, this result shows that maximum kinetic rates are not dependent of the radiation field inside the reactor (the radiative properties being only involved in the definition of radiation losses at the boundaries), or of any state variable for the material phase as the biomass concentration or the kind of mixing (as long as turbulent conditions are guaranteed). Finally, the form of eq. (17) may be easily understood from the consistent energetic analysis presented in appendix of this paper. This appendix indeed established in the general three-dimensional case, and especially in the simplified one-dimensional case, that the maximum spatial radiant light power density absorbed in a photobioreactor (physically light-limited) is given by:

$$<\mathcal{A}>_{\max} \cong \frac{2\alpha}{1+\alpha} a_{light}\ q_{\cap} \quad (19)$$

i.e. only depends on a surface balance in photonic energy onto the reactor and is completely independent of the irradiance profile inside this reactor. So, comparison of eqs. (17) and (19) demonstrates that it is possible to assess the maximum productivities of a PBR just by multiplying $<\mathcal{A}>_{\max}$ with properly defined energetic and quantum yields $\rho_M\ \bar{\phi}$. This is only true at the limit because eq. (17) consistently integrate the fact (in the term $K \ln\left[1+q_{\cap}/K\right]$ instead of $q_{\cap}$) that the energetic yield of photosynthesis strongly decreases when the hemispherical incident PFD $q_{\cap}$ increases. Taking the Taylor expansion of eq. (17) when $q_{\cap}/K$ approaches 0 (or more rigorously the irradiance of compensation $G_C$) is then demonstrative as we obtain directly:

$$<r_X>_{\max} = \rho_M\ \bar{\phi}\ \frac{2\alpha}{1+\alpha} a_{light}\ q_{\cap} = \rho_M\ \bar{\phi} <\mathcal{A}>_{\max} \quad (20)$$

This confirms that a PBR only operates at its maximum thermodynamic efficiency ($\rho_M$) for very low values of the hemispherical incident PFD $q_{\cap}$, completing the energetically consistent analysis of biomass growth rates from radiant light conversion in PBR as sketched in appendix. Conversely, for higher incident photon flux inputs, i.e. in the general case, eq. (20) must be multiplied by an efficiency factor for photosynthesis:

$$E_{\varphi} = \frac{K}{q_{\cap}} \ln\left[1+q_{\cap}/K\right] \quad (21)$$

only related to the hemispherical incident PFD and to the half saturation constant, and limiting the maximum potential biomass productivity whereas the mean radiant light power density increases, according to eq. (19).

## 3.4- Restriction Removal, Generalization and Final Formula

The previous useful formula (eq. 17) has been established with strong restrictive hypotheses both on the geometry and lighting conditions for the PBR and on the considered microorganism; nevertheless, as examined in this paragraph, it can be generalized to almost any generally encountered actual situations of PBR.

### 3.4.1- Other geometries and lighting conditions of PBR

From the general demonstration given in appendix, it is first obvious that the mean spatial radiant light power density absorbed $<\mathcal{A}>$ (eqs 19 and 20) can be easily defined in any given geometry characterized by a specific illuminated area $a_{light}$. For example (see appendix), in case of rectangular PBR illuminated by two opposite sides $a_{light} = 2/L$; in the same way, for a cylindrical vessel radially illuminated, $a_{light} = 4/D$. Moreover, for more complicated situations, it is always feasible to define geometrically $a_{light}$ from eq. (18) with the associated definition of a mean hemispherical incident PFD on the surface $S_{light}$ (see the next paragraph).

On the other hand, the mathematical form of the efficiency factor (eq. 20) comes from the integration of the local kinetic rate (eq. 6) with a simplification for the one-dimensional attenuation for the field of irradiance in the particular case of a rectangular reactor illuminated by one side (eq. 16). If it is easy to demonstrate that it can be generalized to any problem in Cartesian coordinates, its validity is less obvious in other geometries. It is however possible, from a rigorous treatment of the RTE in curvilinear coordinates with an appropriate redefinition of the moving frame of reference (the unit vector $\mathbf{\Omega}$) to prove that the form of eq. (21) applies in any geometry and lighting situation, giving then a general predictive character to eq. (17). This step requiring lengthy and tedious developments remains indeed out of the scope of this paper but an extensive treatment of this problem can be found in other author's publication (*7*).

Finally, the domain of validity of eq. (17) must be still extended to match with many different original and more or less complex concepts of PBR existing in the literature (*5*) such as partially lighting vessels, annular reactors and/or internally lighting PBR. In the former case, it is necessary to introduce a dark volume fraction $f_d$, corresponding to the volume of the PBR which is not illuminated by construction (i.e. a dead volume for the photosensitized reactions). In other cases, the problem is just to properly calculate the specific illuminated area $a_{light}$ as in eq. (18), defining a productivity relative to the working liquid volume (annular) or the total volume (taking into account the internal lighting devices for example) of the reactor and then modifying the final value of $< r_X >_{\max}$ accordingly.

### 3.4.2- Other assumptions on the nature of the hemispherical incident photon flux

Eq. (17) has been established from the assumption of an homogeneous and quasi-collimated hemispherical incident PFD $q_\cap$. Nevertheless, these restricted conditions can be easily enlarged to the more general situations encountered in practice. First, if the incident PFD is not homogeneous on the input surface $S_{light}$, it is always possible to average this flux on the considered surface. This can be done both experimentally, using light cosine sensors with multiple points of measurement or using a chemical actinometer (*7*), or theoretically, from emission model of the artificial sources (*13*), or for example, averaging the diurnal or seasonal variations of the zenith $\varphi$ and azimuth $\theta$ angles in case of solar irradiation.

Many practical cases also deal with non-collimated incident photon fluxes. Here again, a general approach can be proposed, introducing the degree of collimation for the radiation field inside the PBR. It relies on an extension of the two-flux method (historically restricted to collimated or diffuse radiation fields, i.e. intensities independent of the polar angle) recently proposed by the authors (*7*), and consisting in a modulation of the radiant intensities by the $n^{th}$ moment, before integrating over each hemisphere. This degree of collimation *n* enables to take into account all the actual situations with polar dependent intensities between the two limiting previous cases, the collimated field with $n = \infty$, and the diffuse field with $n = 0$. It can be demonstrated (*7*), after a rigorous reformulation of the two-flux method and of the assumption given by eq. (16), or its equivalent in curvilinear coordinates, that a simple correction term $\frac{n+2}{n+1}$ then appears in the general improved eq. (17). This extension is particularly useful, for example, to take in consideration both the direct and diffuse components of any solar irradiation.

### 3.4.3- Eukaryotic microorganisms

For eukaryotic microorganisms of course, the previous assumption used in formulating the kinetic coupling for the establishment of eq. (17) that $r_{X,d} = 0$ is not verified, because the compartmentation (chloroplasts, mitochondria) authorizes the respiration both in darkness or in light. This evidences then, from eq. (2), that a second integral term must be rigorously evaluated for the correct assessment of the mean volumetric growth rate in the PBR. Nevertheless, if the respiration in light remains negligible (as it is often the case), it must be emphasized that eq. (2), or its resulting integral form eq. (17), give again the maximum kinetic performance of any considered PBR cultivating eukaryotic microalgae, because the existence of a negative respiratory term (if the working illuminated fraction $\gamma$ is lower than 1) would lead to a decrease of the volumetric biomass productivity. This demonstrates, as previously stated by the authors with the introduction of the concept of luminostat (*19*), the crucial importance of the radiation field control inside the PBR for optimization purposes. This is particularly true if cultivating eukaryotic microalgae, because in this case, the working illuminated fraction $\gamma$ must be rigorously equal to 1 in order to reach the maximum volumetric productivity for the PBR $< r_X >_{\max}$, as the second integral term in eq. (2) then always vanishes, resulting finally in a perfect validity of eq. (17).

### 3.4.4- Final general formula

Summarizing here all the extensions of the previous restrictions used in building eq. (17) in a final general and useful formula as discussed in the preceding paragraphs leads finally to write the maximum biomass volumetric growth rate in any PBR in the form:

$$< r_X >_{\max} = (1 - f_d)\,\rho_M\,\bar{\phi}\,\frac{2\alpha}{1+\alpha}\,a_{light}\,\frac{K}{\left(\frac{n+2}{n+1}\right)}\ln\left[1 + \frac{\left(\frac{n+2}{n+1}\right)q_{\cap}}{K}\right] \quad (22)$$

or indifferently, with the help of eqs. (20-21) to:

$$< r_X >_{\max} = (1 - f_d)\,\rho_M\,\bar{\phi} < \mathcal{A} >_{\max}\frac{K}{\left(\frac{n+2}{n+1}\right)q_{\cap}}\ln\left[1 + \frac{\left(\frac{n+2}{n+1}\right)q_{\cap}}{K}\right] \quad (23)$$

These simple relations are clearly of considerable interest, taking into account their degree of generality and their predictive character in the sense where all the parameters may be known *a priori*, either from the design of the PBR or from the previous theoretical work of the authors, as explained in this paper. Additionally, even if they have been established in the restricted framework of CSTR, as already discussed above, they appear completely independent of the material phase characteristics (according to the theoretical analysis given in appendix), and particularly of the biomass concentration. They can thus also be used for PFTR as long as their conception has been optimized with a recycle to ensure an optimal functioning in physical light limitation in all their volume. If it is not the case, eqs. (22, 23) can be used to assess the theoretical maximum productivity in a given PFTR for its possible optimization. Moreover, their general form is also available for photoheterotrophic microorganisms, even if the value of $\bar{\phi}$ is more difficult to obtain in this case taking into account the complexity of the photosystem functioning (*19*).

# 4- Experimental Validations

The experimental validations have been performed in eight (very) different, completely stirred PBR, the working liquid volume of which varying between 0.1 and 77 L with dark volume fractions $f_d$ ranging between 0 and 0.48 as described in the M&M section. Numerous kinds of geometries have been investigated with different artificial lighting systems and different ways in mixing the culture medium. For each kind of PBR, the experimental values of productivities have been reported in Table 2. Both batch and continuous cultures have been investigated, and in all cases, the experimental mean volumetric growth rate has been obtained according to the biomass balance onto the well-mixed PBR (defining the so-called residence time for continuous cultures $\tau = V_L / Q_L$ ):

$$< r_X > = \frac{C_X}{\tau} + \frac{d\,C_X}{dt} \quad (24)$$

For calculations of the theoretical maximum productivities with the proposed simple formula, we used the general predictive parameters determined in this paper for $\rho_M$, $\overline{\phi}'$ and $\overline{\phi}$, and specific predictive parameters adapted to *A. platensis* for *Ea*, *Es*, *b* and *K* as determined elsewhere (*22*, *23*). All their values are summarized in Table 1. It must be noticed here that if the radiative properties *Ea*, *Es* and *b* are specific data of a given microorganism, they are only used in eq. (22) for the calculation of the linear scattering modulus $\alpha$. However, calculations of the authors on many different photosynthetic microorganisms experiences that the value of $\alpha$ often ranges between 0.85 and 0.95. Consequently, and considering that the exact calculation of radiative properties is a complicated problem, a mean value of $\alpha = 0.9$ can be used by default with only $\pm 5\%$ of deviation on the final result.

**<u>Table 1</u>**: Values of all the predictive parameters used (general and specific) for applying the simple formula (eq. 22) giving the maximum biomass productivities in PBR.

| **General predictive parameters** | **Specific predictive parameters** (***Arthrospira platensis***) |
|---|---|
| Maximum energetic yield $\rho_M = 0.80 \pm 0.02$ (eq. 7) | Mass absorption coefficient *Ea* = $162 \pm 8$ m$^2$.kg$^{-1}$ |
| Mass quantum yield (nitrate as N-source): | Mass scattering coefficient *Es* = $640 \pm 30$ m$^2$.kg$^{-1}$ |
| $\overline{\phi} = (1.85 \pm 0.10) \times 10^{-9}$ kg$_X$.µmol$^{-1}_{h\nu}$ (eq. 13) | Backscattered fraction $b = 0.030 \pm 2 \times 10^{-3}$ |
| Degree of collimation: $n \to \infty$ (quasi-collimated for all the light sources used at any PFD) | Half saturation constant $K = 90 \pm 5$ µmol.m$^{-2}$.s$^{-1}$ (eq. 15) |

The comparison between the experimental $< r_X >$ and theoretical $< r_X >_{\max}$ biomass productivities calculated by eq. (22) for each kind of PBR and for hemispherical incident PFD $q_{\cap}$ ranging between 30 and 1600 μmol.m$^{-2}$.s$^{-1}$ (PAR) are given in Table 2.

**Table 2** : Comparison between experimental productivities obtained in very different kinds of photobioreactors cultivating *Arthrospira platensis* and the simple formula (eq. 22). The reactors main characteristics and the experimental conditions are given in the M&M section.

| Geometry of the reactor and lighting characteristics | Reactor type and working volume | Operating cultivation condition | Mean incident photon flux density (PAR) ($\mu mol_{h\nu}.m^{-2}.s^{-1}$) | Experimental observed productivity ($kg.m^{-3}.h^{-1}$) | Theoretical maximal productivity given by eq. (22) ($kg.m^{-3}.h^{-1}$) | Deviation (%) |
|---|---|---|---|---|---|---|
| Rectangular, | | batch | 40 | $(1.6 \pm 0.2) \times 10^{-3}$ | $1.8 \times 10^{-3}$ | + 12 |
| lightened by one side | **PBR 1** | batch | 50 | $(2.1 \pm 0.2) \times 10^{-3}$ | $2.4 \times 10^{-3}$ | + 14 |
| $a_{light}$ = 12.5 $m^{-1}$ ($f_d$ = 0) | 4 L | batch | 85 | $(3.2 \pm 0.2) \times 10^{-3}$ | $3.5 \times 10^{-3}$ | + 9 |
| | | batch | 130 | $(2.6 \pm 0.2) \times 10^{-3}$ | $2.8 \times 10^{-3}$ | + 8 |
| | | batch | 260 | $(4.7 \pm 0.4) \times 10^{-3}$ | $4.9 \times 10^{-3}$ | + 4 |
| | | batch | 315 | $(5.0 \pm 0.5) \times 10^{-3}$ | $5.4 \times 10^{-3}$ | + 8 |
| Cylindrical, | | batch | 365 | $(5.3 \pm 0.5) \times 10^{-3}$ | $5.9 \times 10^{-3}$ | + 13 |
| lightened by one side | **PBR 2** | batch | 520 | $(7.1 \pm 0.7) \times 10^{-3}$ | $7.4 \times 10^{-3}$ | + 4 |
| $a_{light}$ = 12.5 $m^{-1}$ ($f_d$ = 0) | 5 L | batch | 575 | $(7.2 \pm 0.7) \times 10^{-3}$ | $7.8 \times 10^{-3}$ | + 8 |
| | | batch | 730 | $(9.5 \pm 0.8) \times 10^{-3}$ | $8.9 \times 10^{-3}$ | - 6 |
| | | batch | 840 | $(1.1 \pm 0.1) \times 10^{-2}$ | $9.6 \times 10^{-3}$ | - 4 |
| | | continuous | 630 | $(8.0 \pm 0.7) \times 10^{-3}$ | $8.3 \times 10^{-3}$ | + 4 |
| | | continuous | 1045 | $(1.2 \pm 0.1) \times 10^{-2}$ | $1.1 \times 10^{-2}$ | - 8 |
| | | continuous | 1570 | $(1.3 \pm 0.1) \times 10^{-2}$ | $1.3 \times 10^{-2}$ | 0 |
| | | batch | 245 | $(1.3 \pm 0.1) \times 10^{-2}$ | $1.4 \times 10^{-2}$ | + 8 |
| Cylindrical, | **PBR 3** | batch | 620 | $(1.9 \pm 0.2) \times 10^{-2}$ | $2.2 \times 10^{-2}$ | + 15 |
| radially lightened | 5 L | batch | 1095 | $(2.7 \pm 0.1) \times 10^{-2}$ | $2.8 \times 10^{-2}$ | + 4 |
| $a_{light}$ = 25 $m^{-1}$ ($f_d$ = 0) | | batch | 1590 | $(3.3 \pm 0.5) \times 10^{-2}$ | $3.2 \times 10^{-2}$ | - 3 |
| | | continuous | 235 | $(1.0 \pm 0.1) \times 10^{-2}$ | $1.1 \times 10^{-2}$ | + 10 |
| Cylindrical, | **PBR 4** | continuous | 365 | $(1.3 \pm 0.1) \times 10^{-2}$ | $1.4 \times 10^{-2}$ | + 7 |
| radially lightened | 7 L | continuous | 625 | $(1.7 \pm 0.2) \times 10^{-2}$ | $1.9 \times 10^{-2}$ | + 12 |
| $a_{light}$ = 40 $m^{-1}$ ($f_d$ = 0.48) | | continuous | 780 | $(1.9 \pm 0.2) \times 10^{-2}$ | $2.1 \times 10^{-2}$ | + 10 |

**Table 2 (continued)** : Comparison between experimental productivities obtained in very different kinds of photobioreactors cultivating *Arthrospira platensis* and the simple formula (eq. 22). The reactors main characteristics and the experimental conditions are given in the M&M section.

| Geometry of the reactor and lighting characteristics | Reactor type and working volume | Operating cultivation condition | Mean incident photon flux density (PAR) ($\mu mol_{h\nu}.m^{-2}.s^{-1}$) | Experimental observed productivity ($kg.m^{-3}.h^{-1}$) | Theoretical maximal productivity given by eq. (22) ($kg.m^{-3}.h^{-1}$) | Deviation (%) |
|---|---|---|---|---|---|---|
| Oblate cylinder, lightened by one side $a_{light}$ = 43.5 $m^{-1}$ ($f_d$ = 0) | **PBR 5** 0.106 L | batch | 65 | $(8.9 \pm 0.1)\times 10^{-3}$ | $9.5\times 10^{-3}$ | + 7 |
| Cylindrical, radially lightened $a_{light}$ = 26.7 $m^{-1}$ ($f_d$ = 0.33) **(experimental results from *31*, *32*)** | **PBR 6** 77 L | batch | 390 | $(1.2 \pm 0.1)\times 10^{-2}$ | $1.3\times 10^{-2}$ | + 8 |
| | | continuous | 525 | $(1.4 \pm 0.2)\times 10^{-2}$ | $1.5\times 10^{-2}$ | + 7 |
| | | continuous | 840 | $(1.7 \pm 0.2)\times 10^{-2}$ | $1.8\times 10^{-2}$ | + 6 |
| Annular and cylindrical, radially lightened $a_{light}$ = 40 $m^{-1}$ ($f_d$ = 0) | **PBR 7** 6 L | batch | 190 | $(2.2 \pm 0.2)\times 10^{-2}$ | $2.0\times 10^{-2}$ | - 10 |
| | | batch | 340 | $(3.1 \pm 0.3)\times 10^{-2}$ | $2.8\times 10^{-2}$ | - 10 |
| | | batch | 530 | $(4.1 \pm 0.3)\times 10^{-2}$ | $3.5\times 10^{-2}$ | - 15 |
| Rectangular, lightened by one side $a_{light}$ = 25 $m^{-1}$ ($f_d$ = 0) **(experimental results from *23*)** | **PBR 8** 0.5 L | batch and continuous | 33 | $(3.3 \pm 0.3)\times 10^{-3}$ | $3.5\times 10^{-3}$ | + 6 |
| | | continuous | 135 | $(1.1 \pm 0.1)\times 10^{-2}$ | $1.0\times 10^{-2}$ | - 10 |

## 5- Discussion and Concluding Remarks

The analysis of the experimental results given in Table 2 and their comparison with the theoretical maximum biomass productivities given by the general formula (eq. 22) is very demonstrative. It clearly establishes that for almost the 30 experiments in completely different conditions (geometry, lighting system, spectral source, mixing…), the theoretical value is in very good agreement (taking into account the experimental accuracy) with the measured biomass volumetric growth rate in physical limitation by light ($\gamma \leq 1$). The maximum deviation observed, as stated in the core of the paper, is $\pm 15\%$ demonstrating the considerable capability of the proposed simple formula (eq. 22) in the particular case of the cyanobacterium *A. platensis*.

This confirms the idea developed in introduction that although the inherent complexity of PBR modeling in the general case, it is possible to easily establish a simple and then useful formula, available in any geometry and operating condition, for the assessment of maximum volumetric productivities in PBR.

The involvements of this formula are of considerable interest for the field of PBR engineering because it rigorously demonstrates the well-known but intuitive idea that the only design parameters in conceiving a good reactor are the specific illuminated area $a_{light}$ (which must be increased as possible) and the dark volume fraction $f_d$ (which must be decreased as possible). The key role played by the hemispherical incident photon flux $q_\cap$ in the PBR performances is also demonstrated: eq. (22) enables to calculate with confidence the possible increase in productivity by increasing $q_\cap$ for a given fixed geometry, but conversely, eq. (21) or its generalization obtained from eq. (23) enables to assess the corresponding loss in thermodynamic efficiency. As illustrative example, it is very simple to calculate that increasing a quasi-collimated incident flux $q_\cap$ from 20 µmol.m$^{-2}$.s$^{-1}$ to 200 µmol.m$^{-2}$.s$^{-1}$ and then to 2000 µmol.m$^{-2}$.s$^{-1}$ (full sun at zenith condition), leads to increase the PBR productivity roughly by 6 and then by 16, but with respective efficiencies (relative to the maximum thermodynamic efficiency for photosynthesis) equal to 0.9 (at 20 µmol.m$^{-2}$.s$^{-1}$), 0.53 and finally 0.14. If the value of the maximum thermodynamic efficiency $\eta_{th}$ at the compensation point for cyanobacteria photosynthesis is taken at 18% (*7*), these values give respectively $\eta_{th}$ = 16%, 9% and 3% for the same respective incident PFD. These estimations are in good agreement with the previous calculations of thermodynamic efficiencies in PBR obtained by the authors with full modeling of light transfer using a sophisticated numerical finite element method (*20*) which confirms, according to the demonstration given in appendix, the energetic consistency of eq. (22). Besides, it must be pointed out also that eq. (22) enables easily to calculate the required increase of specific illuminated area $a_{light}$ if the incident PFD $q_\cap$ is decreased in order to keep constant the kinetic performance of the PBR (and reciprocally). Finally, it must be kept in mind that the proposed formula (eq. 22) gives only the maximum performances of the PBR (CSTR or PFTR operating in condition of physical limitation by light, i.e. using all the incident PFD) but can be thus very useful to discuss losses in volumetric biomass growth rates which could be obtained in case of photoinhibition (too high PFD), in kinetic regime (excess of photons in the PBR) or in light-limited cultures with low values of $\gamma$ for eukaryotic microorganisms (microalgae).

This simple and reliable formula appears then as a robust and practical tool for sizing new original PBR (as internally lighting reactors) or for the comparison of different designs and technologies, independently of the considered microorganism. This last point of course would rigorously required that the experimental results presented and discussed above concern many different photoautotrophic microorganisms, as a proof of the supposed degree of generality claimed along with the paper (specially for the values of the general kinetic parameters $\rho_M$, $\bar{\phi}'$ and $\bar{\phi}$ ). Clearly, the results presented in this study have been focused on the diversity of the PBR technology and size (3 orders of magnitude in volume), and cultivating different microorganisms in these eight different PBR would have represented a considerable amount of work, exceeding the framework of a scientific paper. Another possibility could have been to try a comparison with the numerous experimental results of the literature, but this requires again to know with some confidence the geometrical details of the PBR design ($a_{light}$ and $f_d$) with the corresponding mean values of the PFD $q_\cap$ and to be sure that the data were obtained in conditions of physical limitation by light. This complete information is rarely available in the literature, and except very good results obtained with *Porphyridium cruentum* cultivated in the 100 L ALP2 PBR (*33*, data not shown), it has not been possible to perform this analysis. Nevertheless, a secret wish of the authors would be that the scientific community working in the field of photobioreactor engineering tries to use the proposed simple formula on many different kinds of artificial or solar reactors and different photoautotrophic microorganisms in order to specify its range of validity for the assessment of maximum potential biomass productivities.

## ACKNOWLEDGMENT

The authors thank the European Space Agency (ESA/ESTEC) which partly supported this work through the MELiSSA project.

## NOMENCLATURE

| | |
|---|---|
| $a_{light}$ | Specific illuminated area for any given photobioreactor [m$^{-1}$] |
| $\mathcal{A}$ | Local volumetric radiant power density absorbed [μmol.s$^{-1}$.m$^{-3}$] |
| $b$ | Back-scattered fraction for radiation [dimensionless] |
| c | Speed of light in vacuum, c = 299792458 m.s$^{-1}$ |
| $C_X$ | Biomass concentration [kg.m$^{-3}$ or g.L$^{-1}$] |
| $D$ | Total diameter of a cylindrical photobioreactor [m] |
| $Ea$ | Mass absorption coefficient [m$^2$.kg$^{-1}$] |
| $Es$ | Mass scattering coefficient [m$^2$.kg$^{-1}$] |
| $E_\varphi$ | Efficiency factor [dimensionless] |
| $f_d$ | Design dark volume fraction of any photobioreactor [dimensionless] |
| $G$ | Local spherical irradiance [μmol.s$^{-1}$.m$^{-2}$] |
| h | Planck constant, h = 6.626×10$^{-34}$ J.s |
| $I$ | Specific radiant intensity [μmol.s$^{-1}$.m$^{-2}$] |
| $J_i$ | Molar specific rate for species $i$ [mol.kg$_X$$^{-1}$.s$^{-1}$] |
| k | Boltzmann constant, k = 1.381×10$^{-23}$ J.K$^{-1}$ |
| $K$ | Half saturation constant for photosynthesis [μmol.s$^{-1}$.m$^{-2}$] |
| $L$ | Total length of a rectangular photobioreactor [m] |
| $M_X$ | C-molar mass for the biomass [kg.mol$^{-1}$] |
| $n$ | Degree of collimation for the radiation field [dimensionless] |
| $n_m$ | Refractive index of the liquid medium in the photobioreactor [dimensionless] |
| $p(\mathbf{\Omega},\mathbf{\Omega}')$ | Phase function for scattering [dimensionless] |
| $\mathbf{q}$ | Photon flux density [μmol.s$^{-1}$.m$^{-2}$] |
| $q_\cap$ | Hemispherical incident photon flux density (PFD) in the PAR [μmol.s$^{-1}$.m$^{-2}$] |
| $Q_L$ | Volume liquid flow rate [m$^3$.s$^{-1}$] |
| $Q_P$ | Photosynthetic quotient [dimensionless] |
| $r_X$ | Biomass volumetric growth rate (productivity) [kg.m$^{-3}$.s$^{-1}$ or kg.m$^{-3}$.h$^{-1}$] |
| $S$ | Surface [m$^2$] |
| $S_{light}$ | Illuminated surface of any photobioreactor [m$^2$] |

$t$ Time [s or h]

$T$ Temperature of the photobioreactor [K]

$T_R$ Blackbody radiation temperature [K]

$V$ Volume [$m^3$ or L]

$V_L$ Liquid volume [$m^3$ or L]

$V_\ell$ Illuminated volume inside the photobioreactor [$m^3$ or L]

$V_d$ Dark volume inside the photobioreactor [$m^3$ or L]

$z$ Length [m]

**Greek letters**

$\alpha$ Linear scattering modulus [dimensionless]

$\gamma$ Fraction for working illuminated volume in the photobioreactor [dimensionless]

$\eta_{th}$ Thermodynamic efficiency of the photobioreactor [dimensionless]

$\Theta$ Polar angle [rad]

$\theta$ Azimuth angle [rad]

$\lambda$ Wavelength [m]

$\rho$ Energetic yield for photon conversion [dimensionless]

$\rho_M$ Maximum energetic yield for photon conversion [dimensionless]

$\tau$ Hydraulic residence time [h]

$\upsilon_{ij}$ Stoichiometric coefficient [dimensionless]

$\phi$ Mass quantum yield for the Z-scheme of photosynthesis [$kg_X.\mu mol_{h\nu}^{-1}$]

$\phi'$ Mole quantum yield for the Z-scheme of photosynthesis [$mol_X.\mu mol_{h\nu}^{-1}$]

$\varphi$ Zenith angle [rad]

$\Omega$ Solid angle [rad]

$\mathbf{\Omega}$ Unit directional vector [dimensionless]

**Subscripts**

0 Relative to the input surface of a rectangular photobioreactor

$c$ Relative to the compensation point for photosynthesis

$L$ Relative to the output surface of a rectangular photobioreactor

$\lambda$ Relative to a spectral quantity for the wavelength $\lambda$

max Maximum value for a radiant power density $\mathcal{A}$ or a volumetric productivity $r_X$

**Superscripts**

\+ Relative to the positive hemisphere in defining the radiation field

*ref* Relative to the negative hemisphere in defining the radiation field, i.e. reflected quantity

**Other**

$\overline{X}$ Time or spectral averaging (no possible confusion)

$< X > = \frac{1}{V} \iiint_V X \, dV$ Spatial averaging

**Abbreviations**

CSTR Completely stirred tank reactor

DM Dry mass

PAR Photosynthetically active radiation

PBR Photobioreactor

PFD Photon flux density

PFTR Plug flow tubular reactor

RTE Radiative transfer equation

vvm Gas volume by liquid volume and by minute

## APPENDIX

**Calculation of the maximum spatial radiant light power density $<\mathcal{A}>_{max}$ in a photobioreactor as a proof of the energetic consistency of the proposed relation (eq. 17 or 22)**

It is well documented in the relevant literature (*34*, *35*) that the integration of the RTE (eq. 1 in the paper) over any solid angle $\Omega$:

$$\iint_{4\pi} \nabla \cdot (\mathbf{\Omega} I_\lambda)\, d\Omega = -\iint_{4\pi} (Ea_\lambda + Es_\lambda)\; C_X I_\lambda \; d\Omega + \frac{Es_\lambda C_X}{4\pi} \iint_{4\pi}\iint_{4\pi} I_\lambda \; p_\lambda (\mathbf{\Omega}, \mathbf{\Omega}')\, d\Omega'\, d\Omega$$

leads to the radiant energy balance for the photonic phase of any radiative process. In the particular case of PBR applications, the turbid and participative medium can be considered as non-emitting and this integration gives the following balance at pseudo steady state (*12*, *13*, *24*):

$$-\nabla \cdot \mathbf{q} = \mathcal{A} \quad \text{(A1)}$$

demonstrating that, the elastic scattering being a conservative phenomenon at the local scale, the convergence of the radiative flux $\mathbf{q} = \iint_{\Omega}\int_{\lambda} I_\lambda \cos\Theta \, d\Omega\, d\lambda$ is just balanced by the local radiant light power density absorbed $\mathcal{A}$, defined from the local irradiance $G$ as follows:

$$G = \iint_{\Omega}\int_{\lambda} I_\lambda \, d\Omega\, d\lambda \quad \text{(A2)}$$

$$\mathcal{A} = Ea\, C_X G \quad \text{(A3)}$$

In eq. (A1) the local radiant density absorbed $\mathcal{A}$ may be considered as a sink term for the photonic phase, but eq. (A3) clearly establishes that it is also a source term for the material phase (the volumetric absorption coefficient $EaC_X$ establishing then the link between the photonic field of irradiance $G$ or $\mathbf{q}$ and the actual absorbed energy by the material phase $\mathcal{A}$). As already explained elsewhere (*16*, *39*), the conservation of the total energy, i.e. the first principle of thermodynamics applied on the PBR imposes that this terms be equaled, whatever the method involved for their calculation, from the photonic or the material phases. This conclusion also holds for the mean spatial volumetric radiant light power density absorbed in any PBR and just defined by the volume integral:

$$<\mathcal{A}>=\frac{1}{V}\iiint_V \mathcal{A}\, dV \qquad \text{(A4)}$$

As explained in the core of this paper, this term, corresponding to the total spatial radiant light power density available for the photosensitized reactions can serve as a basis for the assessment of any PBR kinetic performances. Using the definition (A4) together with eq. (A1) and applying the Gauss-Ostrogradsky theorem enables then to write:

$$-\frac{1}{V}\iiint_V \nabla\cdot\mathbf{q}\; dV = <\mathcal{A}> = -\frac{1}{V}\oiint_S \mathbf{q}\cdot d\mathbf{S} \qquad \text{(A5)}$$

On the other hand, eq. (A4) may be combined with eq. (A3) on the material phase to give finally, with eq. (A5) the powerful relation:

$$<\mathcal{A}> = \frac{1}{V}\iiint_V Ea\, C_X\, G\, dV = -\frac{1}{V}\oiint_S \mathbf{q}\cdot d\mathbf{S} \qquad \text{(A6)}$$

This equation of considerable interest is just what we need to prove the consistency of the useful formula (eq. 17 and 22) developed in this paper. First it represents the unique condition for the

verification of the first principle of thermodynamics onto any PBR (*16*, *36*). It is then applicable whatever the geometry of the reactor, the kind of microorganism or the type of functioning (CSTR, PFTR). It clearly demonstrates that it is possible to calculate the energetic and kinetic performances of a PBR from a volume integral of the material phase, or independently, from a surface integral on the photonic phase, i.e. just balancing the input and output mean flux densities at the boundaries. In case of PBR functioning in physical limitation by light ($\gamma \leq 1$), the output flux densities terms vanish (the radiant light energy lost) in the surface integral, demonstrating that the maximum performances are fixed by the specific illuminated area $a_{light} = S_{light}/V$ and the hemispherical incident light flux density $q_{\cap}$, independently of the status of the material phase.

Examining the particular simple case of rectangular PBR illuminated by one side, as in the paper, enables to develop eq. (A6) in the form:

$$< \mathcal{A} >= \frac{1}{L}\int_0^L Ea\, C_X\, G(z)\, dz = \frac{1}{V}\left[S_{light}\left(q_0 - q_L\right)\right] = \frac{1}{L}\left(q_0 - q_L\right) = a_{light}\left(q_0 - q_L\right) \quad \text{(A7)}$$

Additionally, retaining only the case of physical limitation by light ($\gamma \leq 1$) leads to the highest possible value $< \mathcal{A} >_{\max}$ for the integral (A7) in a form corresponding to that obtained in the paper by a different approach:

$$< \mathcal{A} >_{\max} = \frac{1}{L}\int_0^L Ea\, C_X\, G(z)\, dz = a_{light}\, q_0 = a_{light}\left(q_0^{+} - q_0^{ref}\right) = \frac{2\alpha}{1+\alpha}\, a_{light}\, q_{\cap} \quad \text{(A8)}$$

This result rigorously establishes the energetic consistency of the proposed kinetic relations (eq. 17 or 22), if they are interpreted through eq. (19-21). This demonstration is clearly not limited to this particular simple case because of the degree of generality of eq. (A6) already pointed out. If we restricted the field of application of eq. (A6) to PBR functioning in condition of physical limitation by

light which is the only considered case in this paper for the assessment of maximum volumetric biomass productivities, it is evident that any geometry obeys to the more general equation, available in any situation as discussed in the core of the paper:

$$< \mathcal{A} >_{\max} = \max\left(-\frac{1}{V}\oiint_{S} \mathbf{q}\cdot d\,\mathbf{S}\right) = \frac{2\alpha}{1+\alpha} a_{light}\, q_{\cap} \quad \text{(A9)}$$